\documentclass[10pt,english,journal]{IEEEtran}
\usepackage[T1]{fontenc}
\usepackage[latin9]{inputenc}
\usepackage{geometry}
\usepackage{amsmath}
\usepackage{xpatch}
\usepackage{amssymb}
\usepackage{esint}
\usepackage{mathtools}
\usepackage{amsthm}
\usepackage{nicefrac}
\usepackage{bigints}
\usepackage{mathtools}
\usepackage{blkarray, bigstrut}
\usepackage{physics}
\usepackage{calligra}
\usepackage{graphicx}
\usepackage{epstopdf}
\usepackage{dsfont}
\usepackage{array,ragged2e}
\usepackage{enumitem}
\usepackage{etoolbox}
\usepackage{babel}
\usepackage[nopar]{lipsum}
\usepackage{psfrag}
\usepackage[ruled,norelsize]{algorithm2e}
\usepackage{algorithmic}
\usepackage{algorithm2e}
\usepackage{balance}
\usepackage{float}
\usepackage{subcaption}
\SetKw{KwBy}{by}

\makeatletter
\newcommand{\removelatexerror}{\let\@latex@error\@gobble}
\makeatother

\xpatchcmd{\proof}{\hskip\labelsep}{\hskip5\labelsep}{}{}  
\makeatletter
\xpatchcmd{\proof}{\@addpunct{.}}{\@addpunct{:}}{}{}
\makeatother

\renewcommand\[{\begin{equation}}
\renewcommand\]{\end{equation}} 
\usepackage{listings}
\usepackage{fancyvrb}
\usepackage{framed}

\usepackage{courier}
\usepackage[usenames,dvipsnames,table]{xcolor}

\definecolor{dkgreen}{rgb}{0,0.3,0}
\definecolor{gray}{rgb}{0.5,0.5,0.5}

\makeatletter
\newcommand*{\rom}[1]{\expandafter\@slowromancap\romannumeral #1@}
\makeatother

\usepackage{siunitx}
\usepackage{tabu}
\usepackage{booktabs}
\usepackage{multirow}

\usepackage{capt-of}
\usepackage{array}
\usepackage{arydshln}
\DeclareMathOperator*{\argmax}{arg\,max}

\begin{document}
\title{Actor-Critic-Based Learning for Zero-touch Joint Resource and Energy Control in Network Slicing}
\author{Farhad Rezazadeh$^1$, Hatim Chergui$^1$, Loizos Christofi$^2$, and Christos Verikoukis$^1$\\
{\normalsize{} $^1$ Telecommunications Technological Center of Catalonia (CTTC), Barcelona, Spain\\ $^2$ eBOS Technologies Ltd, Lakatamia, Cyprus}\\
{\normalsize{}Contact Emails: \texttt{\{frezazadeh, hchergui, cveri\}@cttc.es, loizos.christofi@ebos.com.cy}}}
\maketitle
\thispagestyle{empty}

\begin{abstract}
To harness the full potential of beyond 5G (B5G) communication systems, zero-touch network slicing (NS) is viewed as a promising fully-automated management and orchestration (MANO) system. This paper proposes a novel knowledge plane (KP)-based MANO framework that accommodates and exploits recent NS technologies and is termed KB5G. Specifically, we deliberate on algorithmic innovation and artificial intelligence (AI) in KB5G. We invoke a continuous model-free deep reinforcement learning (DRL) method to minimize energy consumption and virtual network function (VNF) instantiation cost. We present a novel Actor-Critic-based NS approach to stabilize learning called, \emph{twin-delayed double-Q soft Actor-Critic (TDSAC)} method. The TDSAC enables central unit (CU) to learn continuously to accumulate the knowledge learned in the past to minimize future NS costs. Finally, we present numerical results to showcase the gain of the adopted approach and verify the performance in terms of energy consumption, CPU utilization, and time efficiency.
\end{abstract}


\section{Introduction}
\IEEEPARstart{N}{etwork} slicing is a key enabler for B5G syetems, as it proposes a way of severing the network into different segments by leveraging network softwarization and virtualization technologies such as software-defined networking (SDN) and network function virtualization (NFV). To automate network slicing orchestration, zero-touch network and service management (ZSM) framework reference architecture \cite{ZSM} has been designed by ETSI, but the closed-loop operation of its building blocks is still an open research problem to fulfill an efficient and robust zero-touch management. Indeed, we still need a knowledge plane (KP) that plays the role of a pervasive system within the network by building and maintaining high-level models of what the network is supposed to do, in order to provide services and advice to other elements of the network \cite{KP}-\cite{KDN}. Specifically, the quest of automation and optimal control in dynamic telecommunication environments has aroused intensive research on the applications of DRL. The DRL can provide a promising technique to be incorporated in NS and solve the control and optimization issues.

In this context, \cite{ref4}-\cite{ref5} have presented softwarization approaches in NS. In \cite{vrAIn}, the authors have proposed vrAIn as a dynamic resource controller based on DRL for optimal allocation of computing and radio resources.  Li \emph{et al.} have proposed a deep deterministic policy gradient (DDPG)-based solution to enhance energy efficiency and obtain the optimal power control scheme \cite{ref8}. Correspondingly, \cite{ref9} has proposed a method to learn the optimum solution for demand-aware resource management in C-RAN NS. They have developed a DRL method as GAN-DDQN to handle resource management in NS. In \cite{ref10}, has leveraged advantage Actor-Critic (A2C) and incorporated the long short-term memory (LSTM) to track the user mobility and improve the system utility. More recently, Liu \emph{et al.} have proposed a DRL-based method called, DeepSlicing where they decompose NS problem into a master problem and several slave problems wherein DDPG agents learn the optimal resource allocation policy \cite{ref11}.
In this paper, we present the following contributions:
\begin{itemize}
    \item We propose a KP for B5G NS dubbed KB5G and elaborate on how KP can join the architectural aspects of NS to make a harmonization in a continuous control setting through revisiting ZSM operational closed-loop building blocks. Specifically, we consider CPU and energy consumption control and optimization.
    \item We propose TDSAC as an algorithmic innovation in NS. This stochastic Actor-Critic approach supports continuous state and action spaces in telecommunication while stabilizing the learning procedure and improve time efficiency in B5G. Moreover, it benefits from a model-free approach to underpin dynamism and heterogeneous nature of NS while reducing the need for hyperparameter tuning.
    \item We develop a 5G RAN NS environment called \emph{smartech-v2}. It integrates both CPU and energy consumption simulators with an OpenAI Gym-based standardized interface to ensure reproducible comparison of different DRL algorithms.
\end{itemize}
\begin{figure}[h!]
\centering
\includegraphics[scale=.45]{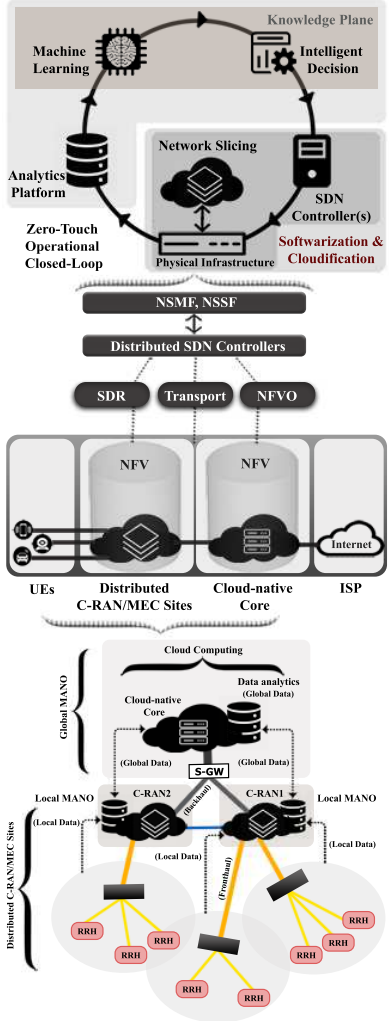}
\caption{Proposed zero-touch KB5G network slicing.}
\vspace{-0.5cm}
\end{figure}

\section{Knowledge-based Beyond 5G (KB5G) Networks}
As depicted in Figure 1, the KP encompasses machine learning (ML) and intelligent decision to handle knowledge discovery, data analysis, and optimization. In what follows we elaborate on the steps in KB5G:\\
\textbf {1) Network Slicing} $\rightarrow$ \textbf{Analytics Platform:} The analytics platform is gathering enough information to offer a complete view of the network and provide current and historical data for feeding learning algorithms. This data is categorized into two types, namely, users' data and operators' data, where both can be either local data or global data. This platform can rely on protocols and functions, such as network configuration protocol (NETCONF) \cite{NETCONF} and network data analytics function (NWDAF) \cite{DA5G}.\\
\textbf {2) Analytics Platform} $\rightarrow$  \textbf{ML} $\rightarrow$ \textbf{Intelligent Decision:} The collected data is utilized by cloud computing platforms to feed learning algorithms for knowledge discovery, data analysis, optimization, and generally manage and control the network to facilitate inferencing. The data analysis represents hidden patterns in big data and can predict and model the future behavior of the network. The KB5G benefits from some indicative abilities of intelligent behavior like learning from experience.\\
\textbf {3) Intelligent Decision} $\rightarrow$ \textbf{SDN Controller(s):} In SDN, the northbound application programming interfaces (APIs) present an abstraction of network functions with a programmable interface to dictate the behavior of the network at the top of the SDN stack. Using declarative languages for the SDN northbound interface and translating intelligent decisions to specific control directives is an open research question yet. The SDN controller receives the declarative primitives through its northbound interface and then renders the intent-driven language into specific imperative control actions \cite{KDN}.\\
\textbf {4) SDN Controller(s)} $\rightarrow$ \textbf{Network:} Due to robustness issues, we consider the distributed control logic \cite{DSDNC}. The distributed SDN consists of multiple interconnected network domains. In NS we have software-defined radio (SDR) for C-RAN, transport controllers, and virtual network function orchestration (NFVO) for cloud-native Core.   The SDN concept can be applied in the KB5G through packet forwarding control protocol (PFCP) instead of OpenFlow-enabled protocol. Indeed, OpenFlow does not support all aspects of quality of service (QoS) issues and it is also packet-based, while PFCP is session-based \cite{LTE}-\cite{EDTV}.

\section{System model}
\vspace{-0.01cm}
Figure 1 shows the considered C-RAN CU-DU split-based network architecture. A total of $N$ access points (APs) are covering $M$ single-antenna users in a downlink setup, and are connected to CU hosting control agents and running as a set of VNFs of the same type. We define $L \in \mathbb{N}$ as the number of slices in the network, and assume that the mobile network operator (MNO) collects the free and unused resources from the tenants and allocate them to the slices in need in a periodic fashion to avoid over-heading. A maximum of $X\in\mathbb{N}$ VNFs can be deployed on top of the cloud, endowed with $Z\,(z=1,\ldots,Z)$ active CPUs having a processing capability of $P_z$ million operations per time slot (MOPTS) \cite{ref14}. Let us denote $\mathbf{h}_m = [h_{1,m},h_{2,m}, ..., h_{N,M}]^H \in \mathbb{C}^{N \times 1} $ as vector of channel gains from the $N$ APs to the $M$ users, where $(\cdot)^H$ is the conjugate transpose and $\mathbb{C}$ represents the complex set. Moreover, we consider the optimal beamforming vector $\mathbf{v}_m = [v_{1,m},v_{2,m}, ..., v_{N,M}]^H \in \mathbb{C}^{N \times 1}$ associated with user $m$ and whose expression is given by \cite{ref15} as $\mathbf{v}_m = \sqrt{p_m} \frac{\left(\mathbf{I}_N + \sum_{j=1}^{M}\frac{1}{\hat{\sigma}^2}\mathbf{h}_j\mathbf{h}_j^{H}\right)^{-1}\mathbf{h}_m}{\norm{ \left(\mathbf{I}_N + \sum_{j=1}^{M}\frac{1}{\hat{\sigma}^2}\mathbf{h}_j\mathbf{h}_j^{H}\right)^{-1}\mathbf{h}_m }}$, where $p_m$ is beamforming power, $\mathbf{I}_N$ denotes the $N\times N$ identity matrix and $\hat{\sigma}^2$ is the noise variance. Therefore we model the received signal $r_m \in \mathbb{C}$ at user $m$ as $r_m = \mathbf{h}_m^H \mathbf{v}_m s_m + \sum_{j \neq m}^{M}\mathbf{h}_m^H \mathbf{v}_j s_j + n_m$, where $s_j \in \mathbb{C}$ is data signal to user $m$ and received noise $n_m$ is the white Gaussian noise with zero mean and variance $\sigma^2$. Let define the following channel model \cite{ref16},
$h_{n,m} = 10^{-L^*(d_{n,m})/20}\sqrt{\vartheta_{n,m}\Theta_{n,m}}g_{n,m}$, where $L^*(d_{n,m})$ denotes the path loss with a distance of $d_{n,m}$. Moreover, $\vartheta_{n,m}$ is the antenna gain, $\Theta_{n,m}$ is the shadowing coefficient and $g_{n,m}$ is the small-scale fading coefficient. Then the achievable rate for user $m$ is given by $R_m=\log(1+\underbrace{\frac{\abs{\mathbf{h}_m^H \mathbf{v}_m}^2}{\sum_{j \neq m}^{M}\abs{\mathbf{h}_m^H \mathbf{v}_j}^2+ \sigma^2 }}_{SINR_m})$, where $SINR$ stands for signal-to-interference-plus-noise ratio. Let define $\mathbf{G}_{n} \in \mathbb{C}^{1 \times N}$ as $\mathbf{G}_{n} = [\underbrace{\hbox{0,...,0,}}_{\hbox{n-1}} 1,0,...,0],\quad n > 0$. Then the power consumption for AP $n$ serving all potential $m$ users can be written as \cite{ref17}, 
$\mathcal{E}_{w}^n = \sum_{m=1}^{M}\mathbf{v}_{m}^H\mathbf{G}_{n}^H\mathbf{G}_{n}\mathbf{v}_{m}$. Note that the circuit power and fronthaul power consumption can be neglected because they are small compared with transmit power. Moreover, we consider energy consumption incurred by the  running processors. The computing resource model follows that in \cite{ref18}. We suppose $\Delta_m$ is a fraction of a CPU core, $\Delta_m = \underbrace{\hat{\theta}R_m+C_0}_{\hbox{baseband}}+\underbrace{\delta\sum_{n = 1}^{N}\Upsilon\abs{\mathbf{v}_{n,m}}}_{\hbox{transmission}}$, where baseband processing refers to coding, modulation and fast Fourier transform (FFT). Furthermore, $\hat{\theta}$ is experimental value, $C_0$ denotes constant complexity for FFT, $\delta > 0$ is slope parameter and $\Upsilon(\cdot)$ denotes the step function. The energy consumed by processor $z$ in Watts is given by $\iota P_{z}^{3}$, where $\iota$ parameter denotes the processor structure \cite{ref19}. We define constant value $\psi$ for VNF deployment and then compute energy consumption in CU with respect to total $\Delta_m$. Therefore, the whole energy consumption in network is given by 
\vspace{-0.2cm}
\begin{equation}
\mathcal{E}_{Net}^{(t)} =
\underbrace{\sum_{z=1}^{Z}\iota P_{z}^{3}+\sum_{x=1}^{X} \psi_{x} }_{\hbox{baseband}}+\underbrace{\sum_{n=1}^{N}\sum_{m=1}^{M}\mathbf{v}_{m}^H\mathbf{G}_{n}^H\mathbf{G}_{n}\mathbf{v}_{m}}_{\hbox{transmission}}
\end{equation}

The objective is to minimize the overall network cost with respect to the incurred computing resources and energy consumption at each decision time step and thereby the continuous model-free DRL optimization is given by
\begin{subequations}
\label{Prob}
\begin{alignat}{2}
&\!\min        &\qquad& \frac{1}{M^{(t)}}(\mathcal{E}_{Net}^{(t)})\\
&\text{subject to} &      & p_m \leq \mathcal{P}_{max},\quad m\in M,\\
&                  &      & SINR_m\geq SINR_{th,l},\quad m\in M,l \in L,\\
&                  &      & \Delta_m\leq \Delta_{th,l},\quad m \in M, l \in L.
\end{alignat}
\end{subequations}
Note that the higher traffic can induce higher costs. We consider the number of users ($M^{(t)}$) at each decision time step to normalize and balance network cost with respect to heavy and low traffic periods. Moreover, $\mathcal{P}_{max}$ is an experimental value while $SINR_{th,l}$ and also $\Delta_{th,l}$ are predefined thresholds for slice $l$.

\section{Problem formulation}
Problem (\ref{Prob}) can be formulated from a Markov decision process (MDP) perspective, where the objective is to achieve lower total costs under user QoS, predefined thresholds, and computing resource constraints. This reflects the correlation between energy consumption and CPU usage, where beamforming power $p_m$ for each user affects SINR that in turn influences computing resource consumption. The MDP can be solved by finding an optimal policy for selecting the best actions with respect to beamforming power and computing resource allocation. Indeed, the MDP is mathematically characterized by a 5-tuple $(S, A, P, \gamma, R)$ where $S$ is the state space, $A$ refers to the action space, $P$ denotes the transition probability from current state $s$ to the next state $s^\prime$, $\gamma$ is the reward discounting hyperparameter, and $R$ stands for the reward function. The state value function for the policy $\pi$ is an explicit measure of how much reward to expect, $
{V}_{\pi}(s)= \mathbb{E}_{\pi}\left[\sum_{n=0}^{\infty}\left(\gamma^{n} R_{t+n+1} | S_t = s \right)\right]
$ and  is defined as action-value function (referred as Q-Function) ${Q}_{\pi}(s,a) = \mathbb{E}_{\pi}\left[\sum_{n=0}^{\infty}\left(\gamma^{n} R_{t+n+1} | S_t = s, A_t = a \right)\right]$. The concerned MPD problem is defined as follows:\\
\textbf{-State space:}
The state space provides input data about possible network configurations for agent via interaction with NS environment parameters. In our scenario, the state transits to the next state at each time step $t$ by $S^{(t)} = \{S_1^{{(t)}}, S_2^{{(t)}}, S_3^{{(t)}}, S_4^{{(t)}}\}$, where $(S_1^{{(t)}})$ is the number of arrival requests for each slice corresponding to each VNF, $(S_2^{{(t)}})$ refers to computing resources allocated to each VNF, $(S_3^{{(t)}})$ shows energy status, $(S_4^{{(t)}})$ refers to number of users being served in each slice.\\
\textbf{-Action space:}
We consider vertical scaling for computing resources consists of either scaling up or down procedure. The CU selects continuous value action with respect to traffic fluctuations to learn how to properly scaling up/down a VNF and thereby according to time step, we have $\mathcal{A}_{CPU}^{(t)} \in \{ o  |   o \in \mathbb{R},-\mathcal{C}_{Net}^{(t)}\leq o \leq \mathcal{C}_{Z}^{(t)}   -\mathcal{C}_{Net}^{(t)}\}$, where $\mathcal{A}_{CPU}^{(t)}$ is vertical scaling action for CPU resources, $\mathcal{C}_{Z}^{(t)}$ is CPU capacity and $\mathcal{C}_{Net}^{(t)}$ denotes the total CPU requirements. Moreover, we assign beamforming power according to SINR constraint, $\mathcal{A}_{P}^{(t)} \in \{ o  |   o \in \mathbb{R}, 0\leq o \leq \mathcal{P}^{(t)}_{max}\}$
the complete continuous multi-action space is given by $\mathcal{A} \triangleq  \mathcal{A}_{CPU}^{(t)} \cup \mathcal{A}_{P}^{(t)}$.\\
\textbf{-Reward:} Due to to guide the agent for learning good results, we define ${\chi}_{T}^{(t)}$ as constraints function that is given by the following piecewise function,
\begin{equation}
\footnotesize
\begin{aligned}
{\chi}_{T}^{(t)}=\left\{
\begin{array}{ll}
\mathbf{0},\quad \mathbf{if}\quad SINR_m \geq SINR_{th,l}\quad\mathbf{and}\quad
\Delta_m \leq \Delta_{th,l}\\
\mathbf{1},\quad \mathbf{otherwise}
                \end{array}
              \right.
\end{aligned}
\end{equation}
Accordingly, we define the penalty function as 
$\varepsilon_{m}^{(t)} = -\varrho_{m} \mathds{1}\left({\chi}_{T}^{(t)} = 1 \right)$, where $\varrho_{m}$ is the penalty coefficient for not fulfilling constraints and $\varrho_{m}^{SINR}>\varrho_{m}^{CPU}$. The objective is maximize the total return $R^{(t)}$,
  \begin{equation}
    R^{(t)} =\frac{\frac{1}{ \frac{1}{M^{(t)}}(\mathcal{E}_{Net}^{(t)})} +  \sum_{m=1}^{M}\varepsilon_{m}^{(t)}}{\hat{\omega}}
  \end{equation}
where $\hat{\omega}$ is a hyperparameter that guarantees $R^{(t)} \in [-1, 1]$. Deep neural network (DNN) uses this return function for training while satisfying the main goal of overall objective function (2).

\section{Twin-delayed double-Q Soft Actor-Critic}
Actor-Critic methods are a combination of policy optimization and Q-Learning. We use $\rho_\pi(s_t)$ and $\rho_\pi(s_t, a_t)$ to denote the state and state-action distribution respectively that induced by policy $\pi$ in NS environment $\pi(a_t|s_t)$. Unlike the DDPG \cite{refnew22} and TD3 \cite{refnew23}, the TDSAC benefits from stochastic policy gradient. The basic idea behind policy-based algorithms is to adjust the parameters $\phi$ of the policy in the direction of the performance gradient $\nabla_{\phi}J(\pi_{\phi})$ concerning the policy gradient theorem \cite{refnew24}. The goal in standard  RL  is to learn a  policy $\pi(a_t,s_t)$ which maximizes the expected sum of rewards. We consider a more general entropy-augmented objective concerning stochastic policies approach where augments the objective with a policy entropy term $\mathcal{H}$ over $\rho_{\pi}(s_t)$. Maximum entropy RL can optimize the expected return and also the entropy of the policy and thereby improves the exploration efficiency of the policy. The objective for finite-horizon MDPs is given by, $J_{\pi} = \mathbb{E} \left[\sum_{i=t}^{T}\gamma^{i-t}[r_i+\alpha\mathcal{H}(\pi(\cdot|s_i))]  \right]$. As we mentioned before, $\gamma$ is the discount factor. The temperature parameter $\alpha$ determines the relative importance of the $\mathcal{H}$ against the reward, thereby handle the stochasticity of the optimal policy. Maximum entropy RL gradually proceeds toward the conventional RL $\alpha \rightarrow 0$.

Let us define entropy-augmented accumulated return or soft return as $G_t = \sum_{i=t}^{T}\gamma^{i-t}[r_i-\alpha\log\pi(a_i|s_i)]$. Then we can define soft Q-value with respect to policy $\pi$ as $Q_{\pi} (s_t, a_t) = \mathbb{E}[r]+\gamma\mathbb{E}[G_{t+1}]$. We use soft policy iteration method for learning optimal maximum entropy policies that alternates between soft policy evaluation and soft policy improvement. In the soft policy iteration, we wish to compute the value of a policy $\pi$ according to the maximum entropy objective \cite{refnew25}, thus the soft Q-value can be learned by applying a  Bellman operator $\mathcal{T^{\pi}}$ under policy $\pi$ repeatedly as, $\mathcal{T^{\pi}}Q_{\pi} (s,a) = \mathbb{E}[r]+\gamma\mathbb{E}[Q_{\pi}(s^{\prime},a^{\prime})-\alpha\log\pi(a^{\prime}|s^{\prime})]$, The optimality and convergence of soft policy iteration have been verified in \cite{refnew26}. The main goal is to find a new policy $\pi_{new}$ that is better than the current policy $\pi_{old}$ and thereby $J_{\pi_{new}} \geq J_{\pi_{old}}$. This particular choice of update can be accomplished by maximizing the entropy-augmented objective $(J_{\pi})$ with respect to soft Q-value, $\pi_{new} = \underset{\pi}{\argmax}\mathbb{E} [Q_{\pi_{old}}(s,a)-\alpha\log\pi(a|s)]$.

Our method (TDSAC) incorporates the following key approaches. The main aim is to stabilize the learning and improve time efficiency while mitigating very high sample complexity and meticulous hyperparameter tuning: $1)$ The (clipped) double Q-learning technique  \cite{refnew23} parameterizes critic networks and critic targets by ${\theta_1}$, ${\theta}_2$ and ${\theta}_1^{\prime}$,${\theta}_2^{\prime}$ respectively. Unlike the TD3 in TDSAC, the next state-actions used in the target come from the current policy ($\phi$) instead of a target policy. $2)$ The target in Q-learning depends on the model's prediction so cannot be considered as a true target. To address this problem, we use another target network instead of using Q-network to calculate the target. $3)$ In TDSAC, the delayed strategy updates the policy, temperature, and target networks less frequently than the value network to estimate the value with a lower variance to have better policy \cite{refnew23}. $4)$ Experience replay enables RL to reuse and also memorize past experiences to solve the catastrophic interference problem. In our method, we store ${({s}_t,{a}_t,{r}_t,{s}_{t+1})}$ to train deep Q-Network and sample random many batches from the experience replay $\beta$ (buffer/queue) as training data. We take a random batch $B$ for all transitions ${({s}_{t_{B}},{a}_{t_{B}},{r}_{t_{B}},{s}_{t_{B}+1})}$.
\begin{algorithm}[ht!]
\caption{TDSAC-based Network slicing}
\scriptsize
\SetAlgoLined
 Initialize actor network $\phi$ and critic networks $\theta_1$, $\theta_2$
 
 Initialize (copy parameters) target networks ${\theta}_1^{\prime}$, ${\theta}_2^{\prime}$

 Initialize learning rate $\ell_{\alpha}, \ell_Q, \ell_{\pi}$
 
 Initialize replay buffer $\beta$
 
 Import custom gym NS environment (`smartech--v2')
 
 \While {t < max\_timesteps}{

  \eIf{t < start\_timesteps}{
   $a$ = env.action\_space.sample()
   }{
   Select action $a\sim\pi_{\phi}(a|s)$
   
  }
  next\_state, reward, done, \_ = env.step($a$)
  
  store the new transition ${({s}_t,{a}_t,{r}_t,{s}_{t+1})}$ into $\beta$
  
   \If{t $\geq$ start\_timesteps}{
   sample batch of transitions ${({s}_{t_{B}},{a}_{t_{B}},{r}_{t_{B}},{s}_{t_{B}+1})}$
   
    $\theta_i\longleftarrow\theta_i-\ell_{Q}\nabla_{\theta_i}J_Q(\theta_i)$, \quad i=1,2 \quad\#Update soft Q-function

   \If{$t \mod freq$}{ 
   $\phi\longleftarrow\phi+\ell_{\pi}\nabla_{\phi}J_{\pi}(\phi)$ \quad \#Update policy weights
   
   $\alpha\longleftarrow\alpha-\ell_{\alpha}\nabla_{\alpha}J(\alpha)$\quad \#Adjust temperature
   
${\theta}_{i}^{\prime}\longleftarrow\tau{\theta}_{i}+(1-\tau){\theta}_{i}^{\prime}$\quad i=1,2\quad \#Update target network
                    }
   }
  \If{done}{
   obs, done = env.reset(), False
   }
  t=t+1
 }
\end{algorithm}

Let us define $Q_{\theta}(s,a)$ and $\pi_{\phi}(a|s)$ as parameterized functions to approximate the soft Q-value and policy, respectively. We consider a pair of soft Q-value functions $(Q_{\theta_1}, Q_{\theta_2})$ and separate target soft Q-value functions $(Q_{\theta^{\prime}_1}, Q_{\theta^{\prime}_2})$. We calculate the update targets of $Q_{\theta_1}$,  $Q_{\theta_2}$ according to $
y= r+\gamma(\underset{i = 1,2}{\min}Q_{\theta^{\prime}_i}(s^{\prime}, a^{\prime})) - \alpha\log\pi_{\phi}(a^{\prime}|s^{\prime}), \quad a^{\prime}\sim\pi_{\phi}
$
we can train soft Q-value by directly minimizing,
\begin{equation}
    J_Q(\theta_i) =\mathbb{E} [(y-Q_{\theta_i}(s,a))^2], \quad i=1,2
\end{equation}

\begin{table*}
\caption{Comparison of hyperparameters tuning in simulation.}
\scriptsize
\centering
\begin{tabular}{@{}lccccccccc@{}}\toprule
\textbf{Architecture} & \textbf{DDPG} &  \textbf{SAC} & \textbf{TD3} & \textbf{our Method (TDSAC)} \\ \midrule
\textbf{Method} & Actor-Critic  & Actor-Critic  & Actor-Critic &  Actor-Critic \\ \hdashline
\textbf{Model Type} & Multilayer perceptron  & Multilayer perceptron  & Multilayer perceptron &  Multilayer perceptron\\ \hdashline

\textbf{Policy Type} & Deterministic &  Stochastic &  Deterministic & Stochastic
\\ \hdashline
\textbf{Policy Evaluation} & TD learning &  Double Q-learning & Clipped double Q-learning &  Clipped double Q-learning \\ \hdashline
\textbf{No. of DNNs}& 4  & 6  & 6  & 5 \\ \hdashline
\textbf{No. of Policy DNNs}& 1  & 1  & 1 &  1 \\ \hdashline
\textbf{No. of Value DNNs}& 1  & 2  & 2  & 2 \\ \hdashline
\textbf{No. of Target DNNs}& 2  & 3  & 3 &  2 \\ \hdashline
\textbf{No. of hidden layers}& 2  & 2  & 2  & 5 \\ \hdashline
\textbf{No. of hidden units/layer}& 200  & 256 &  400/300 & 128 \\ \hdashline
\textbf{No. of Time Steps}& $2e6$ &  $2e6$ &  $2e6$ & $2e6$ \\ \hdashline
\textbf{Batch Size}& 64  & 256  & 100  & 128 \\ \hdashline
\textbf{Optimizer}& ADAM &  ADAM &  ADAM  & ADAM \\ \hdashline
\textbf{ADAM Parameters ($\beta_1, \beta_2$)}& (0.9, 0.999) &  (0.9, 0.999)  &  (0.9, 0.999)   & (0.9, 0.999) \\ \hdashline

\textbf{Nonlinearity}& ReLU  & ReLU & ReLU &  GELU \cite{refGELU28} \\ \hdashline
\textbf{Target Smoothing $(\tau)$}& 0.001 &  0.005  & 0.005  & 0.001 \\ \hdashline
\textbf{Exploration Noise}& $\theta,\sigma=0.15,0.2$  & None &  $\mathcal{N}(0,0.1)$  & None \\ \hdashline

\textbf{Update Interval $(freq)$}& None & None  & 2  & 2 \\ \hdashline
\textbf{Policy Smoothing}& None  & None  & $\epsilon \sim clip(\mathcal{N}(0,0.2), -0.5, 0.5)$ &  None \\ \hdashline
\textbf{Expected Entropy$(\mathcal{H})$}& None & -dim(Action) & None  & -dim(Action) \\ \hdashline
\textbf{Actor Learning Rate}& 0.0001 &  0.0001  & 0.001  & 0.001 \\ \hdashline
\textbf{Critic Learning Rate}& 0.001 & 0.0001 &  0.001 & 0.001 \\ \hdashline

\textbf{Discount Factor}& 0.99  & 0.99  & 0.99  & 0.99 \\ \hdashline

\textbf{Replay Buffer Size}& $1e6$ &  $1e6$ &  $1e6$ &  $1e6$ \\ \hdashline

\bottomrule
\end{tabular}
\vspace{-0.15cm}
\end{table*}

To obtain lower variance estimates, we use the reparameterization trick \cite{refnew25} and reparameterize the policy using a neural network transformation where $a = f_{\phi}(\xi;s)$. Therefore, the policy update gradients with respect to experience replay ($\beta$) is given by
\begin{equation}
\begin{aligned}
\nabla_{\phi}J_{\pi}(\phi) =
\mathbb{E} [-\nabla_{\phi}\alpha\log(\pi_{\phi}(a|s))+(\nabla_aQ_{\theta}(s,a)\\-\alpha\nabla_a\log(\pi_{\phi}(a|s))\nabla_{\phi}f_{\phi}(\xi;s))]
\end{aligned}
\end{equation}

We can update temperature $\alpha$ by minimizing the following objective
\begin{equation}
    J(\alpha)=\mathbb{E}[-\alpha\log\pi_{\phi}(a|s)-\alpha\mathcal{H}]
\end{equation}
To enforce action bounds in algorithms with stochastic policy, we use an unbounded Gaussian as the action distribution \cite{refnew26}. The proposed approach is summarized in Algorithm 1.

\section{Numerical Results}
We use a PyTorch custom environment interfaced through OpenAI Gym as the most famous simulation environment in the DRL community and evaluate our method described in Section \rom{5} against other SoA DRL approaches, namely, TD3 \cite{refnew23}, DDPG \cite{refnew22}, and SAC \cite{refnew26} with a minor change to keep all algorithms consistent. We consider three slices (A, B, and C) with different constraints where the number of new service requests for VNFs follows a distributed homogeneous Poisson process. There exist 20 APs and a maximum of 50 registered subscribers assigned to different slices randomly and the algorithm computes the computing requirements to allocate to the relevant VNF. The dedicated subscribers to Slice-A are less than Slice-B and Slice-C. Table \rom{1} provides a comparison of architectures and hyperparameters while Table \rom{2} presents network parameters. The DNNs structure for the actor-critic networks and target networks are the same. We have set the hyperparameters following extensive experiments \cite{refnewa27}. The evaluation computes every 20000 iterations concerning the average return over the best 3 of 5 episodes. The $start\_timesteps$ denotes the initial time steps for random policy to fill the buffer with enough samples. Moreover, the curves are smoothed for visual clarity in terms of the confidence interval. 

\begin{table}[h!]
\caption{Network parameters in simulation.}
\scriptsize
\centering
\begin{tabular}{@{}lccccccccc@{}}\toprule
\textbf{Network Parameter} & \textbf{Value}\\ \midrule
\textbf{Channel bandwidth} & 10 MHz\\ \hdashline
\textbf{Background noise ($\sigma^2$)} & -102 dBm\\ \hdashline
\textbf{Antenna gain ($\vartheta_{n,m}$)} & 9 dBi\\ \hdashline
\textbf{Log-normal shadowing ($\Theta_{n,m}$)
} & 8 dB\\ \hdashline
\textbf{Small-scale fading distribution ($g_{n,m}$)} & $\mathcal{C}\mathcal{N}$(0, $I$)\\ \hdashline
\textbf{Path-loss at distance $d_{n,m}$ (km)} & 148.1+37.6 $\log_2$($d_{m,n}$) dB\\ \hdashline
\textbf{Distance $d_{m,n}$} & distributed uniformly [0, 600]\\ \hdashline
\textbf{($\iota ,P_{z}$)} & ($10^{-26}, 10 ^ 9$)\\

\bottomrule
\end{tabular}

\end{table}
As shown in Figure 2, the learning curve of TDSAC outperforms all other algorithms in the final performance.
\begin{figure}[h!]
\centering
\includegraphics[scale=.5]{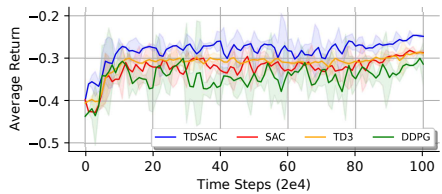}
\caption{Learning curves of the  smartech-v2 network slicing environment and continuous control benchmarks.}
\vspace{-0.15cm}
\end{figure}
Note that the scenario has a big and complex state space. The learning procedures are based on interaction with the NS environment. The NS has different network configurations and parameters (states) and thereby the curves experience high fluctuation during learning. As we mentioned in Sec. \rom{4}, we use a reward-penalty approach for constraints and thresholds in Problem (\ref{Prob}). Indeed, this experimental approach (Eqn. 4) can lead the agent to good results because the problem formulation (\ref{Prob}) is general.

Figure 3 demonstrates the time efficiency of the different algorithms in terms of the wall-clock time consumption on the custom NS environment (smartech-v2). The results show that the TDSAC method yields performance improvement and it has comparable performance to TD3 and lower than SAC.
\begin{figure}[h!]
\centering
\includegraphics[scale=.37]{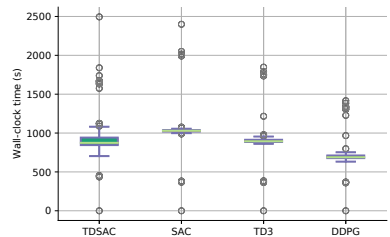}
\caption{Time efficiency comparison of different algorithms on the custom environment (smartech-v2).}
\end{figure}
Note that DDPG uses 4 DNNs (Table  \rom{1}) in its architecture and this results in lower wall-clock time consumption compared to other methods but it has the lowest average return between methods and thereby we should compare wall-clock time with average return.
\begin{figure*}[ht!]
\centering
\subfloat[Energy (Slice A)]{%
      \includegraphics[width=0.19\textwidth]{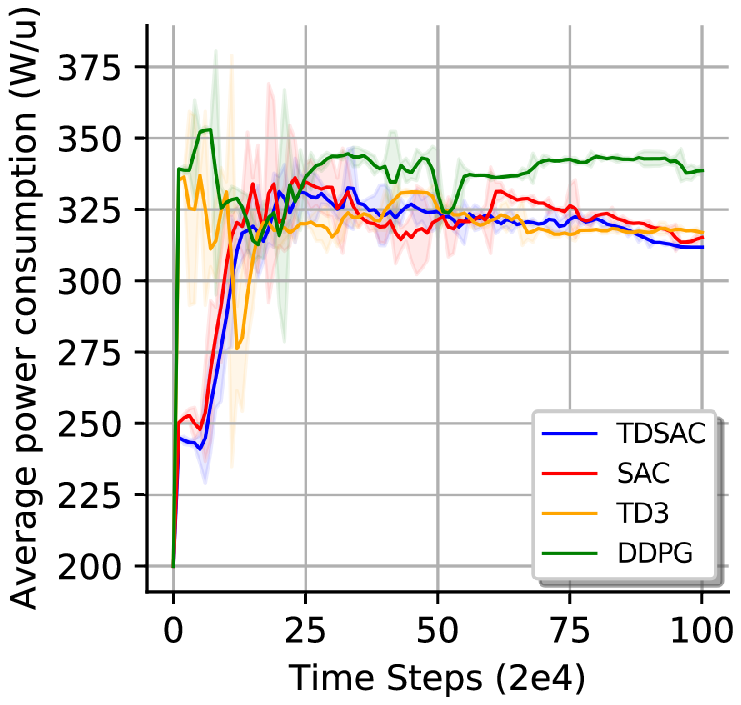}}
       \hfill
\subfloat[Energy (Slice B)]{%
      \includegraphics[width=0.19\textwidth]{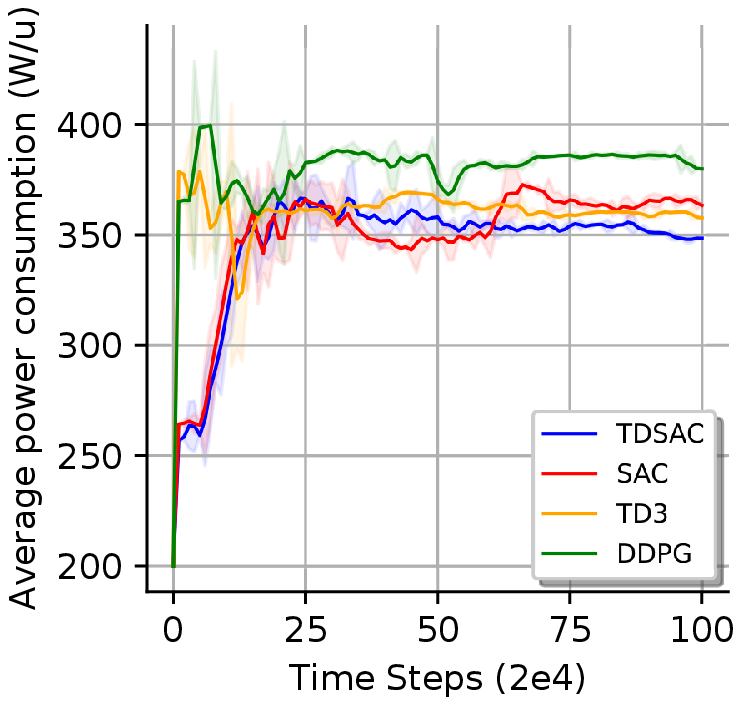}}
      \hfill
\hfill
\subfloat[Energy (Slice C)]{%
      \includegraphics[width=0.19\textwidth]{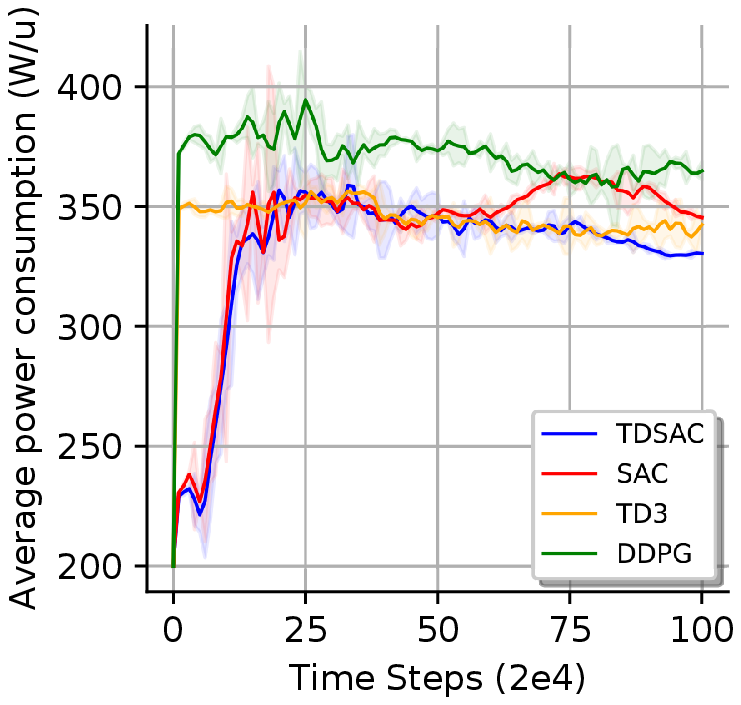}}
       \hfill
\subfloat[Energy (Network)]{%
      \includegraphics[width=0.19\textwidth]{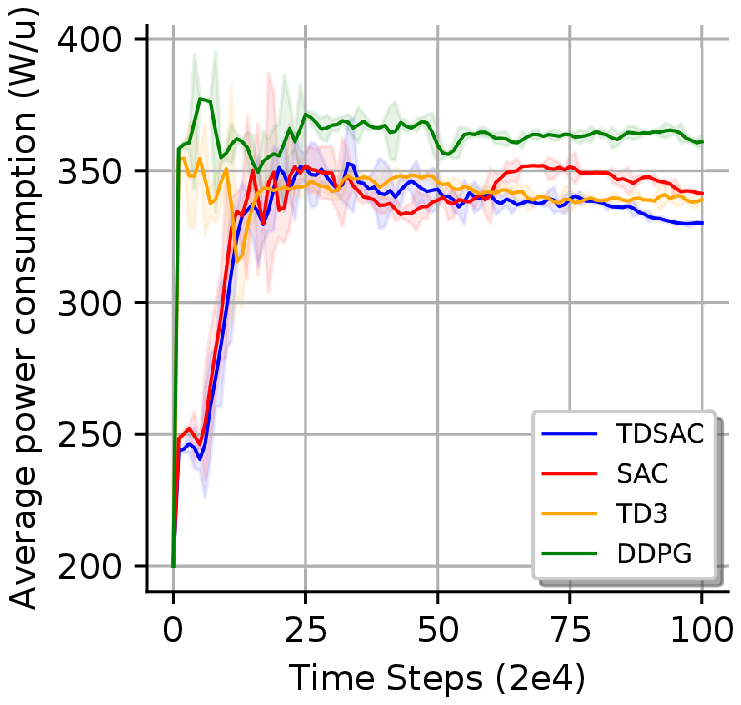}}
      \hfill
\subfloat[CPU utilization]{%
      \includegraphics[width=0.19\textwidth]{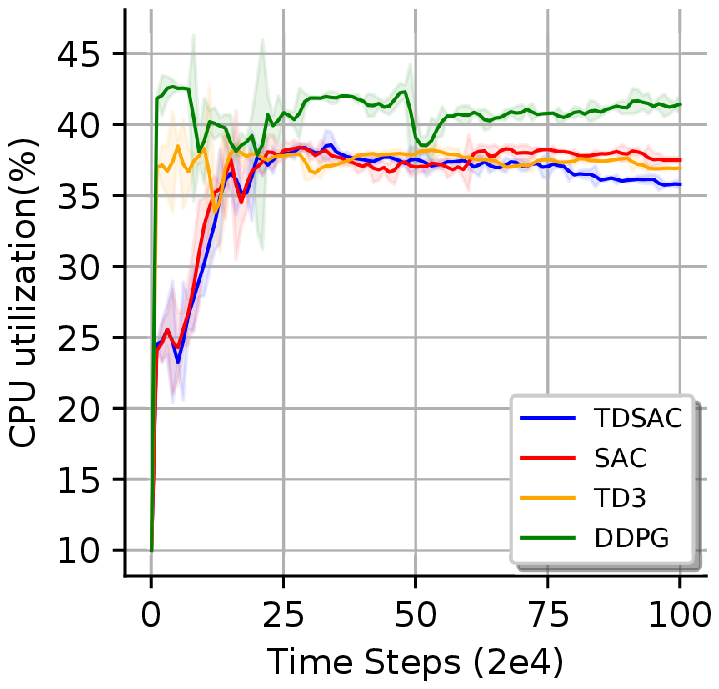}}

\caption{Network performance and costs comparison between TDSAC and other DRL benchmarks. The curves are smoothed for visual clarity. The solid lines demonstrate the mean and the shaded regions correspond to confidence interval over 3 trials.}
\vspace{-0.3cm}
\end{figure*}
All evaluations were run on a single computer with a 3.40 GHz 5 core Intel CPU and evaluation is according to the average per 50 time steps and based on 100 evaluations.

We consider the trade-off between CPU resource usage and energy consumption by defining a cross-layer and correlated cost function (Eqn. 1). Figures 4-(a), 4-(b), and 4-(c) show that the performance of TDSAC is better than other approaches. The agent learns to decrease VNFs instantiation and thereby reduce energy in the baseband part while tuning optimal wireless transmission power. In some scenarios, DDPG cannot learn perfectly because of some issues such as overestimation and lack of stable learning behavior, whereas the TDSAC, TD3, and SAC used the referred techniques (see Section (\rom{5})) to reduce overestimation, stabilize the training, surmount the curse of dimensionality, solve gradient explosion, and mitigate catastrophic forgetting problems. Note that a large part of energy consumption is constant and agent cannot minimize these values.  In Figures 4-(d) and 4-(e), we consider MNO and slices (tenants) as a unified network where slices are isolated and trade-off computing resources with MNO. As shown in Figures 4-(d) and 4-(e), the TDSAC has a better performance compared with other methods. The TDSAC has better resource control between MNO and tenants. We consider CPU utilization efficiency as the ratio of exploited computing resources with respect to the total available CPU for the execution of a VNF.

\section{Conclusion}
To fulfill zero-touch NS, a knowledge-based scheme with an efficient resource provisioning ability should be adopted. We have proposed a KP for B5G NS called, KB5G and elaborated on how KP can solve control and optimization problems in NS. Specifically,  we have deliberated on algorithmic innovation and AI-driven approach and also proposed a continuous model-free DRL method called, TDSAC to minimize energy consumption and VNF instantiation cost. Meanwhile, we have compared the network performance and costs between TDSAC and other DRL benchmarks. We have shown that the proposed solution outperforms other DRL methods.

\section*{Acknowledgement}This work has been supported in part by the research projects 5GSTEPFWD (722429), MonB5G (871780),   5G-SOLUTIONS (856691), AGAUR(2017-SGR-891) and SPOT5G (TEC2017-87456-P).

\end{document}